\documentclass[journal]{IEEEtran}
\usepackage{cite}
\usepackage{amsmath,amssymb,amsfonts}
\usepackage{algorithmic}
\usepackage{graphicx}
\usepackage{textcomp}
\usepackage{caption}
\usepackage{romannum}
\usepackage{color,soul}
\usepackage{ifpdf}
\usepackage{array}
\usepackage{stfloats}
\usepackage{url}
\usepackage{multirow}
\usepackage{supertabular}
\usepackage{subfigure}

\ifCLASSINFOpdf

\begin{document}
%
% paper title
% Titles are generally capitalized except for words such as a, an, and, as,
% at, but, by, for, in, nor, of, on, or, the, to and up, which are usually
% not capitalized unless they are the first or last word of the title.
% Linebreaks \\ can be used within to get better formatting as desired.
% Do not put math or special symbols in the title.
\title{Rational Dynamic Price Model for Demand Response Programs in Modern Distribution Systems\vspace{-0.1cm}}
%\title{Bare Demo of IEEEtran.cls\\ for IEEE Journals}
%
%
% author names and IEEE memberships
% note positions of commas and nonbreaking spaces ( ~ ) LaTeX will not break
% a structure at a ~ so this keeps an author's name from being broken across
% two lines.
% use \thanks{} to gain access to the first footnote area
% a separate \thanks must be used for each paragraph as LaTeX2e's \thanks
% was not built to handle multiple paragraphs
%
\author{Rayees A. Thokar,~\IEEEmembership{Student Member,~IEEE,}
	    Nikhil~Gupta,~\IEEEmembership{Senior Member,~IEEE,}
        K. R.~Niazi,~\IEEEmembership{Senior Member,~IEEE,}
        Anil~Swarnkar,~\IEEEmembership{Senior Member,~IEEE}
        and~Nand K. Meena,~\IEEEmembership{Senior Member,~IEEE}}% <-this % stops a space
    \vspace{-0.1cm}
%\thanks{M. Shell was with the Department
%of Electrical and Computer Engineering, Georgia Institute of Technology, Atlanta,
%GA, 30332 USA e-mail: (see http://www.michaelshell.org/contact.html).}% <-this % stops a space
%\thanks{J. Doe and J. Doe are with Anonymous University.}% <-this % stops a space
%\thanks{Manuscript received April 19, 2005; revised August 26, 2015.}}

% note the % following the last \IEEEmembership and also \thanks - 
% these prevent an unwanted space from occurring between the last author name
% and the end of the author line. i.e., if you had this:
% 
% \author{....lastname \thanks{...} \thanks{...} }
%                     ^------------^------------^----Do not want these spaces!
%
% a space would be appended to the last name and could cause every name on that
% line to be shifted left slightly. This is one of those "LaTeX things". For
% instance, "\textbf{A} \textbf{B}" will typeset as "A B" not "AB". To get
% "AB" then you have to do: "\textbf{A}\textbf{B}"
% \thanks is no different in this regard, so shield the last } of each \thanks
% that ends a line with a % and do not let a space in before the next \thanks.
% Spaces after \IEEEmembership other than the last one are OK (and needed) as
% you are supposed to have spaces between the names. For what it is worth,
% this is a minor point as most people would not even notice if the said evil
% space somehow managed to creep in.

% The paper headers
\markboth{Journal of \LaTeX\ Class Files,~Vol.~14, No.~8, August~2015}%
{Shell \MakeLowercase{\textit{et al.}}: Bare Demo of IEEEtran.cls for IEEE Journals}
% The only time the second header will appear is for the odd numbered pages
% after the title page when using the twoside option.
% 
% *** Note that you probably will NOT want to include the author's ***
% *** name in the headers of peer review papers.                   ***
% You can use \ifCLASSOPTIONpeerreview for conditional compilation here if
% you desire.

% If you want to put a publisher's ID mark on the page you can do it like
% this:
%\IEEEpubid{0000--0000/00\$00.00~\copyright~2015 IEEE}
% Remember, if you use this you must call \IEEEpubidadjcol in the second
% column for its text to clear the IEEEpubid mark.

% use for special paper notices
%\IEEEspecialpapernotice{(Invited Paper)}

% make the title area
\maketitle

% As a general rule, do not put math, special symbols or citations
% in the abstract or keywords.
\begin{abstract}
Demand response (DR) refers to change in electricity consumption pattern of customers during on-peak hours in lieu of financial gains to reduce stress on distribution systems. Existing dynamic price models have not provided adequate success to price-based demand response (PBDR) programs. It happened as these models have raised typical socio-economic problems pertaining to cross-subsidy, free-riders, social inequity, assured profit of utilities, financial gains and comfort of customers, etc.This paper presents a new dynamic price model for PBDR in distribution systems which aims to overcome some of the above mentioned problems of the existing price models. The main aim of the developed price model is to overcome the problems of cross-subsidy and free-riders of the existing price models for widespread acceptance, deployment and efficient utilization of PBDR programs in contemporary distribution systems. Proposed price model generates demand-linked price signal that imposes different price signals to different customers during on-peak hours and remains static otherwise. This makes proposed model a class apart from other existing models.  The novelty of the proposed model lies in the fact that the financial benefits and penalties pertaining to DR are self-adjusted among customers while preserving social equity and profit of the utility. Such an ideology has not been yet addressed in the literature. Detailed investigation of application results on a standard test bench reveals that the proposed model equally cares regarding the interests of both customers and utility. For economic assessment, a comparison of the proposed price model with the existing pricing models is also performed. 
\end{abstract}

% Note that keywords are not normally used for peerreview papers.
\section*{Nomenclature}
\renewcommand{\arraystretch}{0.85}
\begin{supertabular}[l]{p{35pt} p{200pt}}
$P^{\text{MCP}} (t)$ & Market clearing price at $t$th state. \\
$D(i, j, t)$ & Demand of jth customer of $i$th category during $t$th state. \\
$N_{s}$ & Total number of customers in a particular category. \\
$k_{p}(i)$ & Profit factor of utility for $i$th category of customers. \\
$P(i, j, t)$ & Unit price of $j$th customer of $i$th category during the state $t$. \\ 
$\alpha(i, t)$ & Constant of  demand proportionality at $t$th state. \\
$N_{\text{cat}}$ & Total number of customer categories. \\
$N_{\text{on}}$ & Total number of on-peak hours. \\
$N_{\text{off}}$ & Total number of off-peak hours. \\
$T_{\text{on-peak}}$ & Duration of on-peak hours. \\
$T_{\text{off-peak}}$ & Duration of off-peak hours. \\
$t$ & Duration of $t$th state. \\ 
$i$ & Type of customer category. \\
$j$ & Type of customer in a particular category. \\
$T$ & Set of system states. \\
$\Omega_{\text{on}}$ & Set of on-peak hours. \\
$\Omega_{\text{off}}$ & Set of off-peak hours. \\
$\Omega_{\text{cat}}$ & Set of customer categories. \\
$\Omega_{s}$ & Set of customers in a particular category. \\[-0.3cm]
\end{supertabular}

% For peer review papers, you can put extra information on the cover
% page as needed:
% \ifCLASSOPTIONpeerreview
% \begin{center} \bfseries EDICS Category: 3-BBND \end{center}
% \fi
%
% For peerreview papers, this IEEEtran command inserts a page break and
% creates the second title. It will be ignored for other modes.
\IEEEpeerreviewmaketitle

\section{Introduction}

\IEEEPARstart{T}{he} electricity markets around the globe are undergoing fast transformation. In the emerging electricity markets, the dynamic pricing has assumed significant importance. The conventional electricity markets generally offer two types of pricing structures namely, fixed-rate pricing or flat pricing and block-rate pricing. In fixed price structure, the electricity price remains same irrespective of the energy consumption whereas in block-rate pricing structure the per unit rate of electricity changes with the energy consumption slabs. However, the cost of electricity generation is neither fixed nor changes with slabs. The cost of electricity during peak demand hours are much higher than that of during off-peak hours. And so under the conventional pricing structure, the traditional market takes into account the spot pricing risk into the cost-to-serve customers, thereby making them to pay a premium to hedge against financial losses \cite{faruqui2010impact,1198281}. Moreover, cross-subsidy occurs as the cost-causers are not contributing proportionately towards the generation cost, as a result social equity hampers. The generation-cost and consumption-cost gap is dynamic in nature. This leads to emergence of demand response (DR) coupled dynamic pricing in the electricity-market design. The demand response is the change of electricity consumption pattern of the consumers in response to dynamic pricing. The DR may provide several techno-economic and social benefits by changing the consumption pattern of end users in lieu of financial incentives. The annual report of the U.S. DoE and Federal Energy Regulatory Commission (FERC) on demand response (DR) describes in depth the benefits of DR in electricity markets, and carries out a systematic analysis for DR implementation \cite{doe2006benefits,staff2006assessment,federal2007staff,us2009national}. The effective implementation of  DR and larger participation of end customer in DR may play a bigger role in deferring capital investments in generation expansion, to enhance system efficiency and security, to create fairness in retail pricing, to increase electricity market efficiency, protect market players from spot price volatility risks, provide potential environmental benefits, etc.\cite{yang2017opportunities,article,ahn2011optimal,kirschen2005fundamentals,doe2006benefits,staff2006assessment,federal2007staff,us2009national}. 

In literature several price-based DR (PBDR) programs have been reported \cite{hussain2018review,6254817,ghazvini2017demand,palensky2011demand}. In the price-based DR, different types of dynamic pricing models are used some of these may include real time pricing (RTP), time of use (TOU) pricing, critical peak pricing (CPP) and peak time rebate (PTR) \cite{hussain2018review,6254817,ghazvini2017demand,palensky2011demand}. In RTP, the price signal changes in accordance with the whole sale electricity price, usually recieved a day ahead. The TOU offers two or three-level pricing signals during the day thus provide more flexibility to consumers than RTP. The CPP is same as that of the TOU, but it offers very high price signals during certain critical hours/days. The PTR is also similar to CPP, but it provides cash rebate for demand reduction unlike peak time pricing for consumption. The CPP and the PTR price signals have limited operation time as they are employed for only critical situations of short durations. The RTP offers complex pricing structure but has proven to be most efficient and capable to bring many benefits to customers and utilities in the form of reduced bills and peak load reduction \cite{32577,651628,schwarz2002industrial}. However, it requires advanced ICT \cite{ghazvini2017demand} and exposes low consumption consumers to spot pricing risk \cite{faruqui2010impact}, besides more responsiveness of the customers. The TOU is characterized by simplified energy pricing structure thus encourages permanent load shifting without using advanced ICT \cite{faruqui2012time}. The TOU also provide efficient risk management, enhanced market efficiency, reduced billing, improved integration of DERs \cite{wenders1976peak,sheen1995response,yu2006optimal,david1993effect,kirschen2000factoring,celebi2007model,sheen1994time,aalami2008demand,duan2004risk,faruqui2012time}. Moreover, financial incentives offered under TOU pricing signal can act as a feasible alternative to RTP signal \cite{ghazvini2017demand}. A trade-off between risk and reward in dynamic pricing signals from the customers’ perspective is presented in Ref. \cite{FARUQUI201261,faruqui2012time}. The authors illustrated that more the dynamic nature of a pricing signal less will be the risk aversion and more will be the economic efficiency, and vice-versa. Among all dynamic pricing signals, TOU provides compromising results across the spectrum of risk-reward trade-off, therefore, is accepted by majority of the customers. However, TOU pricing is not free from free-riders and cross-subsidy. In Ref. \cite{5677457}, a demand response exchange scheme is employed for the elimination of free-rider problem. Another congenital issue with dynamic pricing is the rebound effect \cite{6345351,7972993,allcott2011rethinking,ericson2009direct} that may occur under highly fluctuating price signals in which customers dramatically delay consumption pattern to maximize financial gains and/or to avoid a peak, but cause a new peak while trying to satisfy the delayed demand after the DR event ends. The pronounced rebound effect may even defeat the purpose of the PBDR program. There are several other key issues in PBDR such as social equity \cite{faruqui2010impact}, ineffective pricing models \cite{faruqui2010impact,berg1998basics,lin2013electricity}, regulatory/market coordination \cite{cutter2012maximizing,yang2017opportunities}, customer behavior \cite{hussain2018review}, customer fears of price volatility \cite{star2010dynamic}, etc. which may overcast the successful implementation of PBDRs. 
\par The existing PBDR programs have not gained much popularity perhaps due to non-active participation of high-consumption customers in DR programs (DRPs) due to their financial flexibility. These high consumption non-participating consumers, enjoying the benefits offered by the active participation of low-income customers, are referred to as free-riders \cite{5156266,5677457,brubaker1975free}. These free-riders may cause substantial distortions of a market \cite{bagnoli1989provision}. Usually, the most prominent free-riders are responsible for cross-subsidy thus hampers social equity and causes unjustified allocation of benefits to other customers. The success of PBDR programs depends upon many factors such as the risk aversion of utilities, better understanding, social equity, fair and transparent allocation of DR benefits to customers, etc. A PBDR program therefore should overcome typical limitations of existing DRPs, viz. free-rider problem, cross-subsidy and rebound effect, while preserving the benefits of customers and social equity and should also offset risk aversion of utilities.    
\par This paper proposes a novel demand contribution based dynamic price model for the successful implementation of PBDR in modern distribution systems which aims to overcome the shortcomings of existing pricing models. The proposed price model generates a demand-linked price signal (DLPS) during on-peak hours of the day.  The DLPS provides dynamic price signal in proportion to customer’s share in peak demand. Proposed price model is unique as it imposes different price signals to different customers during on-peak hours and remains static otherwise. DLPS varies with both time and customers. The proposed modeling intends to be free from cross-subsidy and free-rider problems, mitigate rebound effect, and fairly allocate DR benefits with social equity while preserving the profit of the utilities. Proposed price model is thoroughly implemented and analyzed on a benchmark 33-bus test distribution system and the application results obtained are presented and compared. 
%\par \textcolor{green}{The remaining of the paper is structured as follows. Section \ref{section:PM} presents the proposed methodology. The mathematical modeling of the proposed pricing signal is presented in Section \ref{section:MM}. Simulation results are investigated and analyzed in Section \ref{section:SR}. Finally, the discussions and the concluding remarks are drawn in Section \ref{section:D} and \ref{section:C} respectively}. This para can be removed.... 

\section{Motivations and the Proposed Pricing Model}\label{section:PM}

In the existing PBDR programs, the free-riders avail undue DR benefits and may introduce cross-subsidy by not reducing their demand during on-peak hours, thus severely affects due DR benefits to other customers. The problem worsens in the presence of big free-riders. The root cause of this problem is that the existing dynamic price signals are linked, in some way, with the total demand of the distribution system rather than individual demand. Although the price-based DR is intended to change individual’s demand. The billing of customers has a reflection of both individual’s demand and dynamic price signal generated on account of total demand. There is no discrimination on the basis of participation of individual consumers in DR. The inconsistency exists between the price signal and DR benefits. Due to this inconsistency the DR benefits are not judiciously allocated among customers. Although the customers participating pro-actively in DR are responsible for the reduced price signal. The benefits pass on to all consumers irrespective of their participation level in DR. This ultimately results in cross-subsidy and free-rider problems. Therefore, the price-signal should also take into account the individual demand of the customers for effective implementation of PBDR.
\par
\begin{figure}[!ht]
	\centering
	\includegraphics[width=\linewidth]{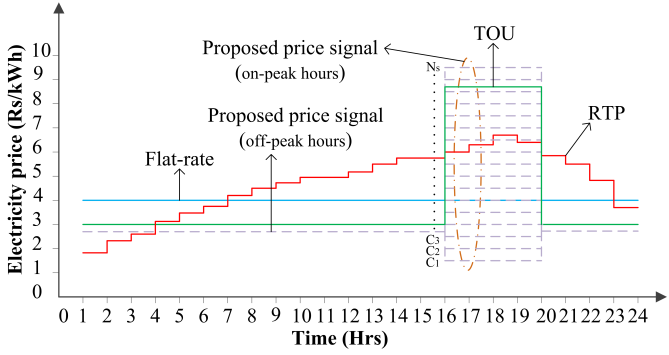}
	\caption{Graphical representation of the proposed pricing structure.}
	\label{Fig:Fig01}
\end{figure}

 Usually, customers prefer TOU on account of its simplicity and less price responsiveness, whereas the utilities prefer RTP owing to risk aversion. This is because the customers can focus on a dynamic price signal, at the most, for few hours while considering their loss of comfort, habits or the mindset. The spot pricing volatility is the main reason behind the preferential choice of RTP by utilities, but it requires almost round-the-clock attention of customers. However, only few hours (on-peak hours) are of significant importance for utilities from techno-economic aspects. Therefore, a dynamic price signal that capture the static nature of pricing during off-peak hours and provides dynamic price signals in proportion to customer’s share in peak demand may be a better alternative to intact benefits of both customers and utility. A graphical representaion of the proposed pricing structure for $N_{s}$ customers along with existing pricing signals is shown in Fig. \ref{Fig:Fig01}. 

\section{Mathematical Modeling of the Proposed Pricing Structure}\label{section:MM}

The mathematical modeling of the proposed dynamic pricing structure is based upon economic balance equation (EBE). EBE equalizes total purchase cost of electricity and the revenue generated for each system state considering demand of individual customer and market clearing price (MCP) while ensuring assured profit of the utility from the purchase and sale of electricity. In this paper, a separate EBE is suggested for each category of customers, and is exclusively modeled for on-peak and off-peak hours. For a particular category, proposed price modeling generates different dynamic price signals to different customers only during the on-peak hours.
 
\subsection{Dynamic Price Signal for On-Peak Hours}

The EBE for the $j$th customer of $i$th category for the state $t$ can be expressed as
\begin{equation}
\scalebox{0.9}[0.9]{$
	\begin{split}
	&(1+{{k}_{p}}(i))P_{{}}^{\text{MCP}}(t)\sum\limits_{j=1}^{{{N}_{s}}}{D(i,j,t)}=\sum\limits_{j=1}^{{{N}_{s}}}{D(i,j,t)}P(i,j,t);\\ & \forall i\in{{\Omega}_{\text{cat}}}, j\in{\Omega_{s}}, t\in{{\Omega}_{\text{on}}} 
	\end{split}
	$}
\label{eqn1}
\end{equation}
 The left and right hand side of the equation \eqref{eqn1} denotes the purchase cost of electricity, including assured profit, and the corresponding revenue generated from electricity sale for the state $t$ respectively.  $k_{p}(i)$ is the profit factor pre-established by the utility for $i$th category of customers and is included in order to ensure profit of the utility. The profit factor may be set different for different categories of customers considering social equity.

For proportional contributory demand, the dynamic price signal can be defined as 
\begin{equation}
P(i,j,t)=\alpha (i,t)\frac{D(i,j,t)}{\sum\limits_{j=1}^{{{N}_{s}}}{D(i,j,t)}}; \forall i\in{{\Omega}_{\text{cat}}}, j\in{\Omega_{s}}, t\in{{\Omega}_{\text{on}}} 
\label{eqn2}
\end{equation}
where, $\alpha (i,t)$ is a constant for demand proportionality that holds during the state $t$ for each customer belonging to $i$th category. Substituting \eqref{eqn2} into \eqref{eqn1} yields
\begin{equation}
\scalebox{0.9}[0.9]{$
	\begin{split}
&\alpha (i,t)=(1+{{k}_{p}}(i))P_{{}}^{\text{MCP}}(t)\frac{{{\left( \sum\limits_{j=1}^{{{N}_{s}}}{D(i,j,t)} \right)}^{2}}}{\sum\limits_{j=1}^{{{N}_{s}}}{{{D}^{2}}(i,j,t)}}; \forall i\in{{\Omega}_{\text{cat}}}, \\ & j\in{\Omega_{s}}, t\in{{\Omega}_{\text{on}}}
\end{split}
$} 
\label{eqn3}
\end{equation}
Using \eqref{eqn3} and \eqref{eqn2}, we get
\begin{equation}
\scalebox{0.9}[0.9]{$
	\begin{split}
&P(i,j,t)=(1+{{k}_{p}}(i))D(i,j,t)\frac{\sum\limits_{j=1}^{{{N}_{s}}}{D(i,j,t)}}{\sum\limits_{j=1}^{{{N}_{s}}}{{{D}^{2}}(i,j,t)}}P_{{}}^{\text{MCP}}(t);\\ &\forall i\in{{\Omega}_{\text{cat}}}, j\in{\Omega_{s}}, t\in{\Omega}_{\text{on}}
\end{split}
$}
\label{eqn4}
\end{equation}
Equation \eqref{eqn4} shows that the dynamic price signal proposed for the $j$th customer of $i$th category during the state $t$ considers profit factor, customer’s demand, demand contribution factor and MCP. Similarly, the price of other category of customers can also be determined. The dynamic price signal defined by \eqref{eqn4} provides economic balance for the customers of $i$th category at $t$th state as below:
\begin{equation}
\scalebox{0.9}[0.9]{$
	\begin{split}
&(1+{{k}_{p}}(i))P_{{}}^{\text{MCP}}(t)\sum\limits_{j=1}^{{{N}_{s}}}{D(i,j,t)}=\sum\limits_{j=1}^{{{N}_{s}}}{D(i,j,t)}\\ &\left(D(i,j,t) (1+{{k}_{p}}(i))\frac{\sum\limits_{j=1}^{{{N}_{s}}}{D(i,j,t)}}{\sum\limits_{j=1}^{{{N}_{s}}}{{{D}^{2}}(i,j,t)}}P_{{}}^{\text{MCP}}(t) \right); \forall i\in{{\Omega}_{\text{cat}}}, j\in{\Omega_{s}}, t\in{\Omega}_{\text{on}}
\end{split}
$}
\label{eqn5}
\end{equation}
Equation \eqref{eqn5} can be extended for all categories of customers for state $t$ and for the interval $T_{\text{on-peak}}$ by the following equations, respectively.
\begin{equation}
\scalebox{0.9}[0.9]{$
	\begin{split}
&P_{{}}^{\text{MCP}}(t)\sum\limits_{i=1}^{{{N}_{\text{cat}}}}{(1+{{k}_{p}}(i))\sum\limits_{j=1}^{{{N}_{s}}}{D(i,j,t)}}=\\ &\sum\limits_{i=1}^{{{N}_{\text{cat}}}}{\sum\limits_{j=1}^{{{N}_{s}}} {D(i,j,t) \left(D(i,j,t)(1+{{k}_{p}}(i))\frac{\sum\limits_{j=1}^{{{N}_{s}}}{D(i,j,t)}}{\sum\limits_{j=1}^{{{N}_{s}}}{{{D}^{2}}(i,j,t)}}P_{{}}^{\text{MCP}}(t) \right)}}; \\ &\forall i\in{{\Omega}_{\text{cat}}}, j\in{\Omega_{s}}, t\in{\Omega}_{\text{on}}
\end{split}
$}
\label{eqn6}
\end{equation}
\begin{equation}
\scalebox{0.9}[0.9]{$
	\begin{split}
&\sum\limits_{t=1}^{{{N}_{\text{on}}}}{P_{{}}^{\text{MCP}}(t)\sum\limits_{i=1}^{{{N}_{\text{cat}}}}{(1+{{k}_{p}}(i))\sum\limits_{j=1}^{{{N}_{s}}}{D(i,j,t)}}}=\\ &\sum\limits_{t=1}^{{{N}_{\text{on}}}}{\sum\limits_{i=1}^{{{N}_{\text{cat}}}}{\sum\limits_{j=1}^{{{N}_{s}}}{D(i,j,t) \left( D(i,j,t)(1+{{k}_{p}}(i))\frac{\sum\limits_{j=1}^{{{N}_{s}}}{D(i,j,t)}}{\sum\limits_{j=1}^{{{N}_{s}}}{{{D}^{2}}(i,j,t)}}P_{{}}^{\text{MCP}}(t) \right)}}}; \\ &\forall i\in{{\Omega}_{\text{cat}}}, j\in{\Omega_{s}}, t\in{\Omega}_{\text{on}}
\end{split}
$}
\label{eqn7}
\end{equation}
\subsubsection{Sensitivity Analysis}
%Sensitivity analysis
 
Using \eqref{eqn4}, let
\begin{equation}
\scalebox{0.9}[0.9]{$
	\begin{split}
&\zeta (i,j,t)=\frac{\partial P(i,j,t)}{\partial D(i,j,t)}=(1+{{k}_{p}}(i))P_{{}}^{\text{MCP}}(t)\frac{\partial }{\partial D(i,j,t)}\\ &\left( \frac{D(i,j,t) \sum\limits_{j=1}^{{{N}_{s}}}{D(i,j,t)}}{\sum\limits_{j=1}^{{{N}_{s}}}{{{D}^{2}}(i,j,t)}} \right); \forall i\in{{\Omega}_{\text{cat}}}, j\in{\Omega_{s}}, t\in{\Omega}_{\text{on}}
\end{split}
$}
\label{eqn8}
\end{equation}
A real distribution system has large number of customers in each category, say residential, industrial, and commercial, etc. or some other consumption-based categories. The equation \eqref{eqn8} then can be safely approximated as
\begin{equation}
\scalebox{0.9}[0.9]{$
	\begin{split}
&\zeta (i,j,t)=(1+{{k}_{p}}(i))P_{{}}^{\text{MCP}}(t)\\ &\frac{\sum\limits_{j=1}^{{{N}_{s}}}{{{D}^{2}}(i,j,t)\sum\limits_{j=1}^{{{N}_{s}}}{D(i,j,t)}-2{{D}^{2}}(i,j,t)\sum\limits_{j=1}^{{{N}_{s}}}{D(i,j,t)}}}{{{\left( \sum\limits_{j=1}^{{{N}_{s}}}{{{D}^{2}}(i,j,t)} \right)}^{2}}}; \\ &\forall i\in{{\Omega}_{\text{cat}}}, j\in{\Omega_{s}}, t\in{\Omega}_{\text{on}}
\end{split}
$}
\label{eqn9}
\end{equation}
\begin{equation}
\scalebox{0.9}[0.9]{$
	\begin{split}
&\Rightarrow \zeta (i,j,t)=(1+{{k}_{p}}(i))P_{{}}^{\text{MCP}}(t)\frac{\sum\limits_{j=1}^{{{N}_{s}}}{D(i,j,t)}}{\sum\limits_{j=1}^{{{N}_{s}}}{{{D}^{2}}(i,j,t)}}; \\ &\forall i\in{{\Omega}_{\text{cat}}}, j\in{\Omega_{s}}, t\in{\Omega}_{\text{on}}
\end{split}
$}
\label{eqn10}
\end{equation}
A comparison of \eqref{eqn3} with \eqref{eqn10} yields
\begin{equation}
\alpha (i,t)=\zeta (i,j,t)\sum\limits_{j=1}^{{{N}_{s}}}{D(i,j,t)}; \forall i\in{{\Omega}_{\text{cat}}}, j\in{\Omega_{s}}, t\in{\Omega}_{\text{on}} 
\label{eqn11}
\end{equation}
Using \eqref{eqn11} and \eqref{eqn2}, it can be shown that 
\begin{equation}
\frac{\partial D(i,j,t)}{D(i,j,t)}=\frac{\partial P(i,j,t)}{P(i,j,t)}; \forall i\in{{\Omega}_{\text{cat}}}, j\in{\Omega_{s}}, t\in{\Omega}_{\text{on}}
\label{eqn12}
\end{equation}
 Equation \eqref{eqn12} reveals that the proposed price model imposes the same percentage change in the price signal as that of the percentage change in customer’s demand, and that too with the same sign.

\subsection{Price Signal for Off-Peak Hours}
The EBE for the off-peak period ($T_{\text{off-peak}}$) is defined as
\begin{equation}
\scalebox{0.9}[0.9]{$
	\begin{split}
&\sum\limits_{t=1}^{{{N}_{\text{off}}}}{P_{{}}^{\text{MCP}}(t)\sum\limits_{i=1}^{{{N}_{\text{cat}}}}{(1+{{k}_{p}}(i))}\sum\limits_{j=1}^{{{N}_{s}}}{D(i,j,t)}}=\\ &\sum\limits_{t=1}^{{{N}_{\text{off}}}}{\sum\limits_{i=1}^{{{N}_{\text{cat}}}}{P(i)\sum\limits_{j=1}^{{{N}_{s}}}{D(i,j,t)}}}; \forall i\in{{\Omega}_{\text{cat}}}, j\in{\Omega_{s}}, t\in{\Omega}_{\text{off}}
\end{split}
$}
\label{eqn13}
\end{equation}
The cost of electricity purchase, including assured profit, and revenue from its sale during the off-peak period are respectively shown in the left and right part of \eqref{eqn13}. In the off-peak period, the unit price is kept same for all the customers belonging to given category. The unit price to all $i$th category customers during the state $t$ is given by
\begin{equation}
\scalebox{0.9}[0.9]{$
	\begin{split}
P(i)=(1+{{k}_{p}}(i))\frac{\sum\limits_{t=1}^{{{N}_{\text{off}}}}{\sum\limits_{j=1}^{{{N}_{s}}}{D(i,j,t)P_{{}}^{\text{MCP}}(t)}}}{\sum\limits_{t=1}^{{{N}_{\text{off}}}}{\sum\limits_{j=1}^{{{N}_{s}}}{D(i,j,t)}}};\forall i\in{{\Omega}_{\text{cat}}}, j\in{\Omega_{s}}, t\in{\Omega}_{\text{off}}
\end{split}
$}
\label{eqn14}
\end{equation}
Proposed model delivers price signals given by \eqref{eqn4} and \eqref{eqn14} which are based upon independent balancing of EBEs defined by \eqref{eqn1} and \eqref{eqn13}, respectively. Therefore, ensures profit of utility over the day and also addresses social equity by taking suitable values of  $k_{p}(i)$ for different categories of customers. The simulation results achieved while employing the proposed price model for on-peak hours for a single state $t$ are discussed in the coming section.

\section{Simulation Results}\label{section:SR}
In this section, the proposed pricing model is investigated on 12.66 kV, 33-bus benchmark test distribution system \cite{baran1989network}. For this system, the nominal active and reactive load demands are 3.715 MW and 2.30 MVar, respectively. The power losses and minimum node voltage at nominal loading are 202.67 kW and 0.9131p.u. respectively.
\subsection{Case Study Data and Validation of the Proposed Pricing Model}\label{section:A1} 
\begin{figure}[!ht]
\centering
\includegraphics[width=\linewidth,height=1.8in]{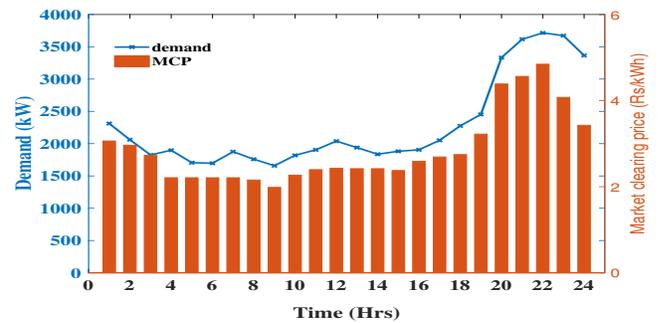}
\caption{Load demand and MCP considered for the sample day.}
\label{Fig:Fig1}
\end{figure}

A sample day is considered to be composed of 24 states each of the duration 1 Hr. 
The period from 18:00 Hrs to 23:00 Hrs is assumed to be peak demand period and the remaining period is taken as off-peak hours, as in \cite{IEX}. The system buses are arbitrarily divided into three categories, viz. residential (N2--N15), commercial (N16--N22, N30--N33) and industrial (N23--N29) as suggested in \cite{8272222}. A total 177 customers are assumed which are divided as 23 industrial (I1--I23), 106 residential (R1--R106) and 48 commercial (C1--C48) customers. The detail of the customers' loading is given in Appendix \ref{section:A}. It is also assumed that all customers are participating in DRP. The load demand and MCP profiles considered for the sample day are taken from \cite{IEX} and shown in Fig. \ref{Fig:Fig1}. The profit factor $k_{p}$ for a particular customer’s category is decided by the utility keeping in view its assured profit. The profit factor for industrial, residential and commercial customers in the present study are assumed as 1.01, 1.03 and 1.02, respectively. In other words, a profit of 1\%, 3\% and 2\% are expected on the sale of energy to industrial, residential and commercial customers respectively.
\begin{table*}[ht]
	\begin{center}
	\caption{Results of the proposed dynamic price signal for base case}
	\label{Table:dynamic price signal for base case}
	\begin{tabular}{|c|c|c|c|c|c|c|c|c|}
		\hline
		\textbf{Customer No.} & \textbf{LD} & \textbf{UP} & \textbf{Customer No.} & \textbf{LD} & \textbf{UP} & \textbf{Customer No.} & \textbf{LD} & \textbf{UP} \\ \hline
		I1                    & 40          & 3.22        & I9                    & 30          & 2.41        & I17                   & 30          & 2.41        \\ \hline
		I2                    & 50          & 4.02        & I10                   & 42          & 3.38        & I18                   & 60          & 4.82        \\ \hline
		I3                    & 46          & 3.70        & I11                   & 54          & 4.34        & I19                   & 25          & 2.01        \\ \hline
		I4                    & 54          & 4.34        & I12                   & 62          & 4.99        & I20                   & 35          & 2.81        \\ \hline
		I5                    & 65          & 5.23        & I13                   & 68          & 5.47        & I21                   & 40          & 3.22        \\ \hline
		I6                    & 75          & 6.03        & I14                   & 74          & 5.95        & I22                   & 38          & 3.06        \\ \hline
		I7                    & 88          & 7.07        & I15                   & 90          & 7.24        & I23                   & 42          & 3.38        \\ \hline
		I8                    & 92          & 7.40        & I16                   & 30          & 2.41        & \multicolumn{3}{c|}{\tiny{\textbf{LD: load demand}; \textbf{UP: unit price}}}                             \\ \hline
	\end{tabular} \\
\end{center}
\end{table*}
\begin{figure}[!ht]
	\centering
	\includegraphics[width=\linewidth]{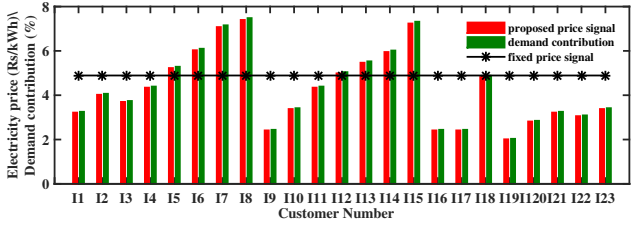}
	\caption{Comparing fixed price and proposed dynamic price considering demand contribution of customers for the given state.}
	\label{Fig:Fig2}
\end{figure}
\par The proposed pricing model is now applied to determine the unit price of all customers for a given state. The analysis is carried for all the system states considering all customers. However, for the ease of analysis the investigation results of industrial customers for the state 22:00 Hr. are presented. The MCP of this state is Rs 4.85. With 1\% profit the fixed unit selling price is Rs 4.89. Table \ref{Table:dynamic price signal for base case} shows the load demand (LD) and unit price (UP) of all the 23 industrial customers, i.e. I1--I23. For this LD, the UP offered to different customers using the proposed dynamic price signal is presented in the table. It can be observed that the UP varies from Rs 2.01 to Rs 7.40, against the fixed price of Rs 4.89. Using MCP and UP the purchase cost of energy and revenue earned by utility are found to be Rs. 5960.12 and Rs. 6019.73, respectively. The profit of the utility is thus Rs 59.60, which is 1\% of the purchase cost.  If the UP is uniformly kept at Rs 4.89 i.e. fixed price signal generated by MCP for all the customers, the revenue received is found to be Rs. 6019.73 again. This validates that the profit of the utility from the purchase and sale of energy remains unaltered using fixed price or proposed dynamic price signal.
\par
\begin{figure}
	\centering
	\includegraphics[width=\linewidth]{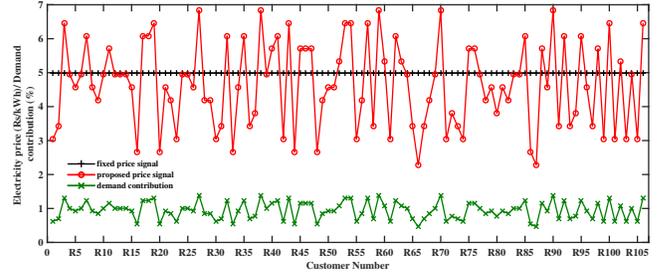}
	\caption{Comparing fixed price and proposed dynamic price considering demand contribution of residential customers for the given state.}
	\label{Fig:Fig2.1}
\end{figure}

\begin{figure}
	\centering
	\includegraphics[width=\linewidth]{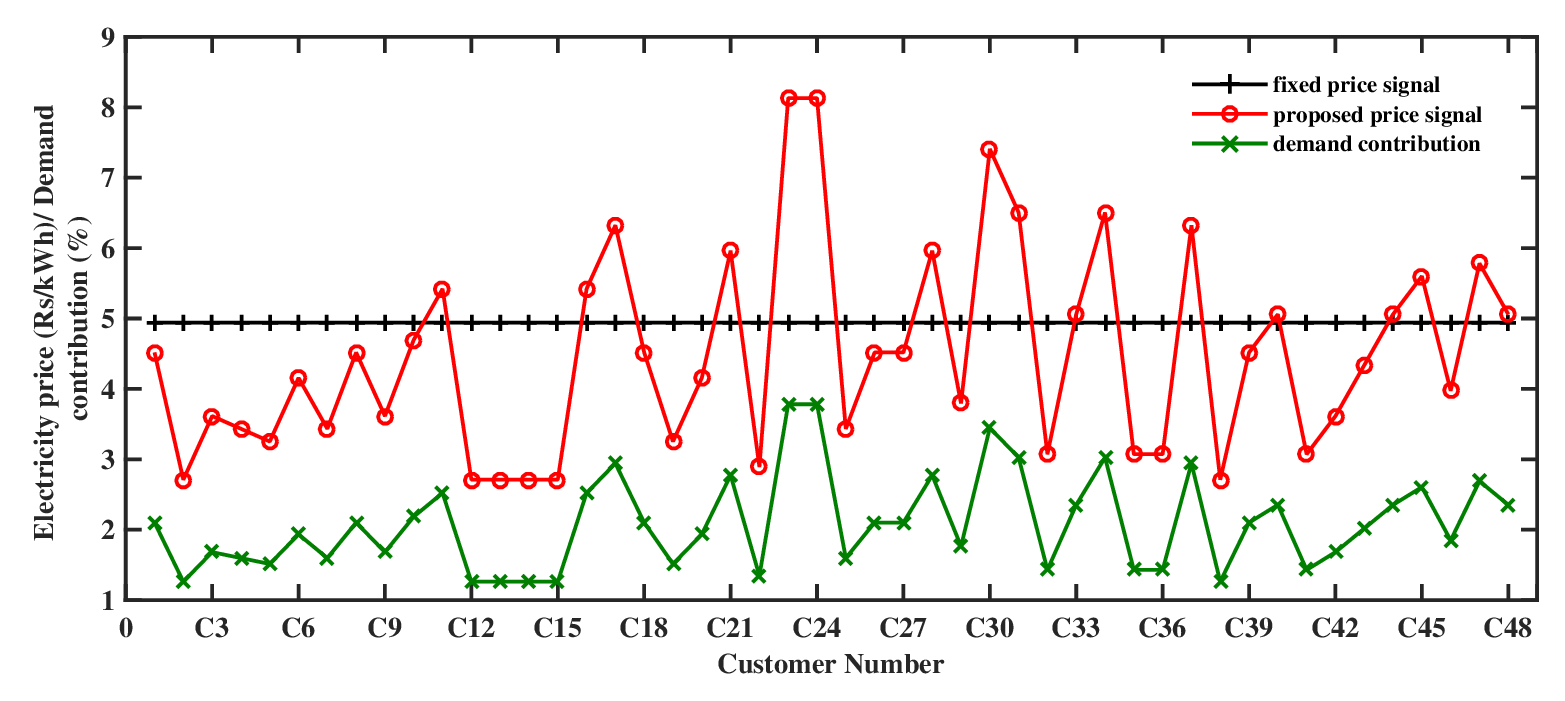}
	\caption{Comparing fixed price and proposed dynamic price considering demand contribution of commercial customers for the given state.}
	\label{Fig:Fig2.2}
\end{figure}

 The UP under fixed and proposed dynamic price signals is compared in Fig. \ref{Fig:Fig2}. It can be observed that the UP using proposed price signal is higher to fixed price only for fewer customers (I5--I8, I12--I15) having demand more than  the approximate mean demand of 60 kW. The figure also shows the demand contribution (DC) of all customers. It can be observed that the proposed price signal for each customer is almost proportional to demand contribution of the customer in the peak demand. Therefore, customers with relatively low demand are getting benefited in comparison to flat pricing system. However, the remaining relatively high demand customers have to pay more as per their increased contribution towards peak demand. This shows that the proposed method mitigates the problem of cross-subsidy. Further, from Fig. \ref{Fig:Fig2.1} and Fig. \ref{Fig:Fig2.2}, it can be observed that the UP of individual customers belonging to residential and commercial category, respectively, is also changing proportionally on the basis of their contribution in the peak demand.
\begin{figure}[!ht]
	\centering
	\includegraphics[width=\linewidth]{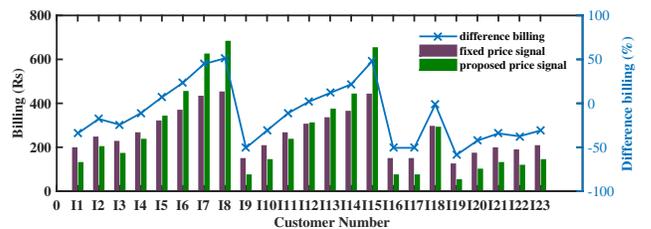}
	\caption{Comparison of billing with fixed price and proposed dynamic price signals for the given state.}
	\label{Fig:Fig3}
\end{figure}   
\par The hourly billings of customers with fixed and proposed price signals are compared in Fig. \ref{Fig:Fig3}. The hourly billing refers to the billing during the given state and will be simply considered as billing throughout the sections \ref{section:A1} \& \ref{section:B1}. The figure shows that by employing the proposed price signal, the billing depends upon the deviation of demand of customer from the mean demand of the group. The billing is more if the demand is above the mean demand and will be less if it is below the mean demand, and will remain same if the demand is equal to the mean demand. Higher the deviation from the mean demand, more is the variation in billing. The figure shows that low demand customers can have as much as $60\%$ reduced billing whereas high demand customers may have up to $52\%$ increased billing as compared to fixed price signal. The analysis shows that the customers get benefited in billing using proposed dynamic price signal only if their contributions in peak demand are below mean demand of the group, and vice-versa. Thus the proposed pricing system can abolish cross-subsidy without affecting profit of the utility from the sale and purchase of energy. This may greatly relieve the utilities as they do not need to bother about the rebates in billing on the basis of their participation in DRP. 
%\textcolor{green}{This exercise and the numerical results obtained satisfactorily validates the functionality and data assumptions of the proposed method as expected. After validation of the proposed pricing model with fixed price model, it is extended for comparison with other existing pricing models used in DRPs like RTP and TOU}.
\subsection{Detailed Scenario-based Investigation of the Proposed Pricing Model}\label{section:B1}

\par In order to further validate the robustness of the proposed pricing system, following scenarios are hypothetically considered assuming that the demand of some of the customers would change.
\begin{enumerate}
	\item \textit{Scenario--1 (S1): Demand of two customers is increased by $10\%$}
	\item \textit{Scenario--2 (S2): Demand of two customers is decreased by $10\%$}
	\item \textit{Scenario--3 (S3): Demand of one customer is increased by $10\%$and that of another is decreased by $10\%$}
	\item \textit{Scenario--4 (S4): Non--uniform increase or decrease in demand} ($-10\%$ to $+10\%$) 
\end{enumerate}
The customer number I8 and I19 have maximum and minimum demand respectively. In order to investigate the effect of proposed pricing model on maximum demand and minimum demand customers, the demand of these customers is only changed in scenarios S1 to S3. The application results of the proposed pricing for scenarios, S1--S3, are summarized in Table \ref{Table:Scenarios S1, S2 and S3}. In the table, the change in load demand ($\Delta$LD) from base value, corresponding change in unit price ($\Delta$UP), change in demand contribution ($\Delta$DC) and change in billing ($\Delta$B) for all the customers are presented. In scenario S1, the customers I8 and I19 have increased their demand by 10\%, assuming that the demand of remaining customers remains unchanged. In other words, customers 18 and I19 are not willing to participate in DR. Therefore, these non-willing participants (customers) may be treated as free-riders. The table shows that there is an increase in UP of free-riders by 8.29\% with consequent increase in billing by 19.11\%. This shows that the free-rider problem can be eliminated using the proposed method as they are suitably penalized in the proposed pricing model. The table also shows evenly reduction in billing of the remaining customers by 1.56\%. This marginal reduction may be justified as these customers restrain themselves from increasing their demands.
\begin{table*}[ht]
	\centering 
	\caption{Results obtained for Scenarios S1, S2 and S3}
	\label{Table:Scenarios S1, S2 and S3}
	\begin{tabular}{cccccccccc}
		\hline
		\multicolumn{1}{|c|}{\multirow{2}{*}{}} & \multicolumn{3}{c|}{\textbf{S1}}                                                                        & \multicolumn{3}{c|}{\textbf{S2}}                                                                        & \multicolumn{3}{c|}{\textbf{S3}}                                                                        \\ \cline{2-10} 
		\multicolumn{1}{|c|}{}                  & \multicolumn{1}{c|}{\textbf{I8}} & \multicolumn{1}{c|}{\textbf{I19}} & \multicolumn{1}{c|}{\textbf{RC}} & \multicolumn{1}{c|}{\textbf{I8}} & \multicolumn{1}{c|}{\textbf{I19}} & \multicolumn{1}{c|}{\textbf{RC}} & \multicolumn{1}{c|}{\textbf{I8}} & \multicolumn{1}{c|}{\textbf{I19}} & \multicolumn{1}{c|}{\textbf{RC}} \\ \hline
		\multicolumn{1}{|c|}{\textbf{$\triangle$LD (\%)}}       & \multicolumn{1}{c|}{+10}         & \multicolumn{1}{c|}{+10}          & \multicolumn{1}{c|}{0}           & \multicolumn{1}{c|}{-10}         & \multicolumn{1}{c|}{-10}          & \multicolumn{1}{c|}{0}           & \multicolumn{1}{c|}{-10}         & \multicolumn{1}{c|}{+10}          & \multicolumn{1}{c|}{0}           \\ \hline
		\multicolumn{1}{|c|}{\textbf{$\triangle$UP (\%)}}       & \multicolumn{1}{c|}{+8.29}       & \multicolumn{1}{c|}{+8.29}        & \multicolumn{1}{c|}{-1.56}       & \multicolumn{1}{c|}{-8.75}       & \multicolumn{1}{c|}{-8.75}        & \multicolumn{1}{c|}{+1.39}       & \multicolumn{1}{c|}{-8.69}       & \multicolumn{1}{c|}{+11.60}       & \multicolumn{1}{c|}{+1.46}       \\ \hline
		\multicolumn{1}{|c|}{\textbf{$\triangle$B (\%)}}        & \multicolumn{1}{c|}{+19.11}      & \multicolumn{1}{c|}{+19.11}       & \multicolumn{1}{c|}{-1.56}       & \multicolumn{1}{c|}{-17.88}      & \multicolumn{1}{c|}{-17.88}       & \multicolumn{1}{c|}{+1.39}       & \multicolumn{1}{c|}{-17.82}      & \multicolumn{1}{c|}{+22.76}       & \multicolumn{1}{c|}{+1.46}       \\ \hline
		\multicolumn{1}{|c|}{\textbf{$\triangle$DC (\%)}}       & \multicolumn{1}{c|}{+8.96}       & \multicolumn{1}{c|}{+8.96}        & \multicolumn{1}{c|}{-0.94}       & \multicolumn{1}{c|}{-9.13}       & \multicolumn{1}{c|}{-9.13}        & \multicolumn{1}{c|}{+0.96}       & \multicolumn{1}{c|}{-9.51}       & \multicolumn{1}{c|}{+10.60}       & \multicolumn{1}{c|}{+0.55}       \\ \hline
		\multicolumn{1}{|c|}{\textbf{PC (Rs)}}       & \multicolumn{3}{c|}{6016.82}                                                                            & \multicolumn{3}{c|}{5903.43}                                                                            & \multicolumn{3}{c|}{5927.66}                                                                            \\ \hline
		\multicolumn{1}{|c|}{\textbf{R (Rs)}}        & \multicolumn{3}{c|}{6076.99}                                                                            & \multicolumn{3}{c|}{5962.47}                                                                            & \multicolumn{3}{c|}{5986.94}                                                                            \\ \hline
		\multicolumn{1}{|c|}{\textbf{P (\%)}}        & \multicolumn{3}{c|}{1.00}                                                                               & \multicolumn{3}{c|}{1.00}                                                                               & \multicolumn{3}{c|}{1.00}                                                                               \\ \hline
		\multicolumn{10}{l}{\begin{tabular}[c]{@{}l@{}}$\triangle$LD: change in load demand; $\triangle$UP: change in unit price; $\triangle$B: change in billing; RC: remaining customers;\\  PC: purchase cost; R: revenue; P: profit\end{tabular}}                           
	\end{tabular}
\end{table*}
\par In scenario S2, the customers I8 and I19 are assumed to decrease their demand by 10\%. This shows their pro-active participation in DR. From the table, it may be observed that the UP of these customers has decreased by 9.0\%. In this case, the remaining customers become free-riders as they are not actively contributing in DR. Consequently, it may be observed that the UP of these free-riders is increased by 1.39\%.                                                                                                                                
\par In Scenario S3, the customer I8 decreases demand by 10\% and customer I19 increases demand by 10\%. This means customer I8 is actively participating in DR whereas customer I19 is a free-rider in existing pricing model who has increased its demand. From the table, it may be observed that the UP of I8 has decreased by 8.69\% and that of I19 has increased by 11.60\%. The UP of remaining customers has marginally increased due to their indifference towards DRP. This shows that in the proposed pricing model there remains no free-rider.
\par It can be observed that the penalty imposed in billing to I19 is about 23\% in Scenario S3 and the rebate is about 18\% in Scenario S2. It happened as $\Delta$DC of I19 is decreased by about 9\% in S2, but is increased by about 11\% in S3. The table also shows that the profit of the utility remains intact at 1\% for all the scenarios (S1--S3). However, the revenue of the utility affects marginally. It happened because the proposed methodology ensures profit of utility as a fixed percentage of purchase cost of energy which depends upon total demand.
\begin{figure}[!ht]
	\centering
	\includegraphics[width=\linewidth]{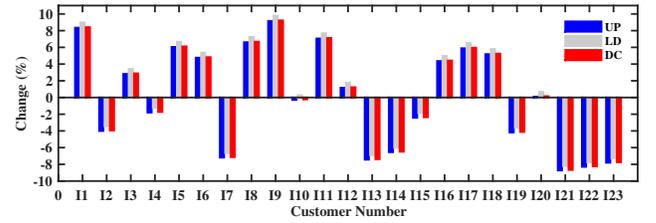}
	\caption{Comparison of change in LD, UP and DC in Scenario S4.}
	\label{Fig:Fig4}
\end{figure}
\par Finally, considering the most practical scenario S4 where the demand of all the customers is assumed to be randomly varied in the range of $\pm10\%$. The percentage change in LD, UP and DC obtained are presented for comparison in Fig. \ref{Fig:Fig4}. A general observation can be drawn from the figure that the proposed method causes almost equal change in UP against the  change in LD of a customer, and this change in UP is very close to the change in DC of the customer, and the sign of change in LD, UP and DC always remains same. However, very minor exceptions may occur if the magnitude of change in LD is around $1\%$ which is acceptable from practical considerations, as in case of the customers I10 and I20.

\begin{table}[h]
	\centering
	\caption{Financial comparison of S4 with base case}
	\label{Table:comparison of S4 with base case}
	\begin{tabular}{c c c c}
		\hline
		\textbf{Particular(s)}      & \textbf{Base case} & \textbf{Scenario S4} & \textbf{Difference} \\ \hline
		Demand (kW)        & 1230               & 1236.42              & 6.42                \\ 
		Purchase cost (Rs) & 5960.12            & 5991.21              & 31.09               \\ 
		Revenue (Rs)       & 6019.73            & 6051.12              & 31.40               \\ 
		Profit (Rs)        & 59.60              & 59.91                & 0.31                \\ \hline
	\end{tabular}
\end{table}

\par The Table \ref{Table:comparison of S4 with base case} shows the purchase cost, revenue received and profit incurred to utility in S4 and their comparison with base case. It can be seen from the table that the profit is increased by Rs 0.31 in S4. It happened on account of additional demand of 6.42 kW. The table shows that additional purchase cost is Rs 31.09 for this demand and the corresponding additional revenue received is Rs 31.40. This increases the profit in S4 by Rs 0.31 which is $1\%$ of the additional purchase cost. This is interesting as the sale price of the additional demand of 6.42 kW under the proposed dynamic price signal is unknown, yet the percentage profit remains intact. It implies that the generated price signals adjust themselves for the given profit despite being governed by the individuals’ contribution towards the peak demand.
\par The above exercise (\ref{section:A1} \& \ref{section:B1}) and the numerical results obtained so far satisfactorily validates the functionality and data assumptions of the proposed model as expected. After validation of the proposed pricing model with fixed price model, it is extended for comparison with other existing pricing models used in DRPs.
\subsection{Comparative Analysis of the Proposed Pricing Model }
\begin{figure}[h]
	\centering
	\includegraphics[width=3.2in,height=1.8in]{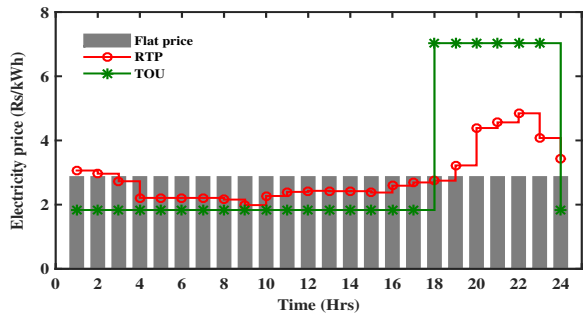}
	\caption{Pricing signals considered for the sample day.}
	\label{Fig:Fig5}
\end{figure}

A comparison of the proposed price signal with the existing price signals like flat, RTP and TOU is performed to analyze and investigate the impact of DR on billing of different categories of customers such as industrial, residential and commercial during peak hours (18:00 Hrs to 23:00 Hrs). For this purpose, MCP profile shown in Fig. \ref{Fig:Fig1} is taken as RTP signal while as flat and TOU price signals are generated from this RTP signal in accordance with Ref. \cite{FARUQUI201261}. The generated pricing signals like flat, RTP and TOU and load demand profile considered for the 24 hours are shown in Fig. \ref{Fig:Fig5} and Fig. \ref{Fig:Fig1} respectively. The comparative analysis can also be extended by considering other forms of dynamic price signals like CPP, PTR etc.
\begin{figure}[!ht]
	\centering
	\subfigure[]{\includegraphics[width=2.7in,height=1.5in]{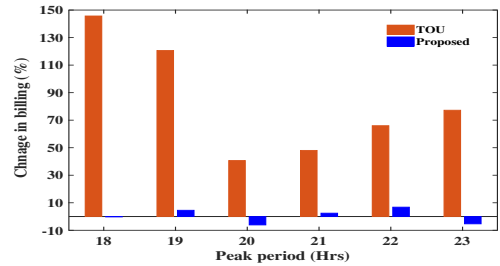}}	
	\subfigure[]{\includegraphics[width=2.7in,height=1.5in]{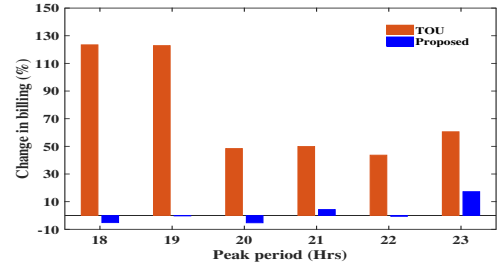}}		
	\subfigure[]{\includegraphics[width=2.7in,height=1.5in]{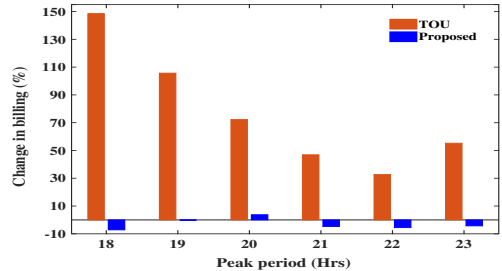}}
	\caption{Graphical representation of state-wise percentage change in billing of (a) industrial customers, (b) residential customers and (c) commercial customers during peak hours.}
	\label{fig:Fig6}
\end{figure}
  
\par In this section, only economic assessment, in terms of billing of customers, is considered to test the effectiveness of the proposed price signal. The technical assessment is not under the scope of this study. Therefore, the aim of DR here is to shed or curtail the demand in place of shifting. For this analysis, it is assumed that all the categories of customers like residential, commercial and industrial are actively participating in demand response i.e., all the customers are reducing their demand in the range from 0-15\% randomly during peak hours. The comparison of collective customer billing over the sample day and during peak hours for all the categories of customers are shown in Table \ref{tab:Table4} and Table \ref{tab:Table5} respectively. From the Table \ref{tab:Table4}, it can be observed that the aggregate billing by the proposed price model gives nearly the same results as obtained by RTP price model. It is important to recall that RTP is based on MCP and that is why it is considered to be most accurate. It may also be observed from the table that aggregate billing obtained by flat rate price signal is appreciably less than RTP which is a loss to the utility. On the other hand, the aggregate billing obtained by TOU pricing signal is much higher than that obtained by RTP. The higher billing offered by TOU is a loss to the customers and therefore it will have lesser acceptability to the customers. These results verify that the proposed price model despite mitigating cross-subsidy and free-rider problems does not affect overall billing in terms of revenue. The comparison of collective customer billing during peak hours for all the categories of customers are shown in Table \ref{tab:Table5}. From the table, it may be observed that TOU price model is charging exorbitantly higher from the customer during peak hours as compared to RTP pricing model. This exorbitantly higher billing has no justification. However, the proposed price model gives nearly the same results as obtained by RTP pricing model.

\begin{table}[]
	\centering
	\caption{Comparison of aggregate billing over the 24 hours \label{tab:Table4}}
	\begin{tabular}{|c|c|c|c|c|}
		\hline
		\multicolumn{1}{|l|}{\multirow{2}{*}{\textbf{Price signal}}} & \multicolumn{3}{c|}{\textbf{Billing (Rs)}}                                                                                                                                                                 & \multicolumn{1}{l|}{\multirow{2}{*}{\textbf{Total (Rs)}}} \\ \cline{2-4}
		\multicolumn{1}{|l|}{}                                       & \begin{tabular}[c]{@{}c@{}}Industrial   \\ customers\end{tabular} & \begin{tabular}[c]{@{}c@{}}Residential   \\ customers\end{tabular} & \begin{tabular}[c]{@{}c@{}}Commercial   \\ customers\end{tabular} & \multicolumn{1}{l|}{}                                     \\ \hline
		Flat                                                         & 51667.58                                                          & 54397.98                                                           & 49987.34                                                           & 156052.9                                                  \\ \hline
		RTP                                                          & 54395.74                                                          & 58713.36                                                           & 53720.32                                                          & 166829.4                                                  \\ \hline
		TOU                                                          & 63171.28                                                          & 66365.98                                                           & 60928.84                                                          & 190466.1                                                 \\ \hline
		Proposed                                                     & 54425.49                                                          & 59383.41                                                           & 52962.47                                                          & 166771.4                                                  \\ \hline
	\end{tabular}
\end{table}
\begin{table}[]
	\centering
	\caption{Comparison of aggregate billing over the peak hours \label{tab:Table5}}
	\begin{tabular}{|c|c|c|c|c|}
		\hline
		\multicolumn{1}{|l|}{\multirow{2}{*}{\textbf{Price signal}}} & \multicolumn{3}{c|}{\textbf{Billing (Rs)}}                                                                                                                                                                 & \multicolumn{1}{l|}{\multirow{2}{*}{\textbf{Total (Rs)}}} \\ \cline{2-4}
		\multicolumn{1}{|l|}{}                                       & \begin{tabular}[c]{@{}c@{}}Industrial   \\ customers\end{tabular} & \begin{tabular}[c]{@{}c@{}}Residential   \\ customers\end{tabular} & \begin{tabular}[c]{@{}c@{}}Commercial   \\ customers\end{tabular} & \multicolumn{1}{l|}{}                                     \\ \hline
		Flat                                                         & 18041.93                                                          & 18995.37                                                           & 17455.2                                                           & 54492.50                                                  \\ \hline
		RTP                                                          & 24180.33                                                          & 26271.26                                                           & 24198.09                                                          & 74649.68                                                  \\ \hline
		TOU                                                          & 41400.83                                                          & 42991.18                                                           & 39657.83                                                          & 124049.84                                                 \\ \hline
		Proposed                                                     & 24210.08                                                          & 26941.31                                                           & 23440.25                                                          & 74591.64                                                  \\ \hline
	\end{tabular}
\end{table}
\begin{table*}[b]
	\centering
	\caption{Comparison of state-wise billing of all the customers over the peak hours \label{tab:Table6}}
	\begin{tabular}{|c|c|c|c|c|c|c|c|}
		\hline
		\multirow{3}{*}{\textbf{\begin{tabular}[c]{@{}c@{}}Customer\\ category\end{tabular}}} & \multirow{3}{*}{\textbf{Price signal}} & \multicolumn{6}{c|}{\textbf{Billing (Rs)}}                \\ \cline{3-8} 
		&                                        & \multicolumn{6}{c|}{\textbf{Peak hours}}                  \\ \cline{3-8} 
		&                                        & 18      & 19      & 20      & 21      & 22      & 23      \\ \hline
		\multirow{4}{*}{\begin{tabular}[c]{@{}c@{}}Industrial\\    customers\end{tabular}}    & Flat                                   & 2153.09 & 2321.20 & 3154.46 & 3421.92 & 3515.67 & 3475.61 \\ \cline{2-8} 
		& RTP                                    & 1929.75 & 2380.41 & 4816.69 & 4995.68 & 5217.18 & 4840.63 \\ \cline{2-8} 
		& TOU                                    & 4741.07 & 5251.57 & 6775.33 & 7395.10 & 8659.09 & 8578.68 \\ \cline{2-8} 
		& Proposed                               & 1924.63 & 2488.36 & 4523.20 & 5117.46 & 5572.54 & 4583.90 \\ \hline
		\multirow{4}{*}{\begin{tabular}[c]{@{}c@{}}Residential \\ customers\end{tabular}}     & Flat                                   & 2266.87 & 2443.86 & 3321.16 & 3602.76 & 3701.45 & 3659.28 \\ \cline{2-8} 
		& RTP                                    & 2197.67 & 2618.22 & 5128.18 & 5221.79 & 6029.29 & 5076.13 \\ \cline{2-8} 
		& TOU                                    & 4909.78 & 5834.38 & 7610.34 & 7827.19 & 8660.66 & 8148.84 \\ \cline{2-8} 
		& Proposed                               & 2084.19 & 2610.09 & 4857.19 & 5445.94 & 5993.89 & 5950.01 \\ \hline
		\multirow{4}{*}{\begin{tabular}[c]{@{}c@{}}Commercial\\ customers\end{tabular}}       & Flat                                   & 2083.07 & 2245.71 & 3051.88 & 3310.64 & 3401.33 & 3362.58 \\ \cline{2-8} 
		& RTP                                    & 2022.23 & 2372.34 & 4233.49 & 5200.21 & 5669.29 & 4700.53 \\ \cline{2-8} 
		& TOU                                    & 5025.88 & 4879.22 & 7291.99 & 7638.25 & 7526.04 & 7296.46 \\ \cline{2-8} 
		& Proposed                               & 1877.38 & 2368.11 & 4391.65 & 4950.39 & 5353.95 & 4498.77 \\ \hline
	\end{tabular}
\end{table*}
\par The Table \ref{tab:Table6} shows the state-wise comparison of customer billing during peak hours. The percentage change in billing of industrial, residential and commercial customers with reference to the RTP signal during peak hours 
%is presented in Table \ref{tab:Table7} and the graphical representation of this percentage change in billing 
is shown in Fig. \ref{fig:Fig6}(a)--\ref{fig:Fig6}(c). 
%In this table the ‘–’ sign indicates percentage reduction in billing while as ‘+’ sign denotes percentage increase in billing of customers and the same follows in Fig. \ref{fig:Fig6}(a)--\ref{fig:Fig6}(c).
In these figures, the change in billing below horizontal axis indicates percentage reduction in billing while as the change in billing above horizontal axis denotes percentage increase in billing of customers.  
Among existing dynamic pricing signals, RTP signal is known to offer potentially highest reward but with more risk. On the Contrary, the TOU signal offers the least potential reward at the lowest risk \cite{faruqui2010impact,faruqui2011dynamic,faruqui2012time}. However, from Fig. \ref{fig:Fig6}(a)--\ref{fig:Fig6}(c) it can be observed that the proposed price signal can offer more reward (in terms of bill savings) most of the times as compared to RTP signal with lowest risk as it remains static during off-peak hours just like TOU price signal. The proposed price model shows gradual enhancement in reward when moving from high consumption to low consumption customers. This can bring a sense of motivation and eagerness in low consumption customers (which are usually low income customers) to actively participate in DRPs unlike existing price models. Also, it is interesting to note that the proposed price signal offers lowest billing for customers during 18:00 Hrs in all the three categories, which is evident from Fig. \ref{fig:Fig6}(a)--\ref{fig:Fig6}(c). Therefore, the proposed price signal is best suitable, promising, least-risky and economically efficient for both the risk-averse as well as risk-taking customers among the dynamic pricing options.

 \section{Conclusion}\label{section:C} 
A demand contribution based dynamic price model is presented for successful implementation of PBDR in the distribution systems. The price signal is designed in such a way that the unit price varies proportionately with the individual’s concurrent demand during on-peak hours, but remains static otherwise. The proposed price model is class apart than other existing counterparts as it imposes different dynamic price signals to different customers during the same time period. 
\par The proposed dynamic price model has been thoroughly investigated on benchmark 33-bus test distribution system under different operating scenario. The application results obtained are also compared with existing dynamic pricing models. The simulation results reveal that the proposed price model varies billing of customers in accordance to the change in their demand. Detailed investigation of the obtained results reveals that the proposed model can overcome chronic problems of DRPs such as cross-subsidy and free-riders while the customers get fair and transparent allocation of financial benefits. The proposed dynamic pricing model suitably rewards the customers who are actively participating and penalizes the customers who do not actively participate in dynamic pricing programs due to their financial flexibility. The comparative analysis shows that the proposed price signal can offer more reward, in terms of bill savings, as compared to other dynamic price signals with lowest risk. The analysis shows gradual enhancement in procuring reward by the customers when moving from high consumption to low consumption customers. This can bring a sense of motivation and interest in low consumption customers to actively participate in DRPs unlike existing dynamic price models. In addition, the proposed model not only preserves assured profit of the utilities but also fully relieved them to deliver financial benefits of DR to the participating customers. The chances of rebound effect may be less using the developed model because it imposes diverse price differentials among the customers. Moreover, the major chunk of the customers gets benefited in electricity billing thus enhances fair chances in DR participation. Proposed model attempts to provide win-win situation to both the customers and the utility. The model is quite unique and innovative, however, may lack public acceptance initially unless supported by intense DR education programs, as in existing models. It is highly simplified, straight forward, and effective but requires advanced ICT infrastructure for implementation. The proposed price model preserves assured percentage profit of utility. The present work can also be used for emergency demand response programs.

\appendices
\section*{Appendix}
\label{section:A}
\renewcommand{\thetable}{A-\arabic{table}}
\setcounter{table}{0}
The detail of the customers' loading is given in Table \ref{Table:Hourly load data}.
\begin{table*}[h]
	\centering
	\caption{Hourly load data assumed for the individual customers of the system under study}
	\label{Table:Hourly load data}
	\begin{tabular}{cccccccccccccccc}
		\hline
		\multicolumn{1}{|c|}{\textbf{CT}} & \multicolumn{1}{c|}{\textbf{LD}} & \multicolumn{1}{c|}{\textbf{CT}} & \multicolumn{1}{c|}{\textbf{LD}} & \multicolumn{1}{c|}{\textbf{CT}} & \multicolumn{1}{c|}{\textbf{LD}} & \multicolumn{1}{c|}{\textbf{CT}} & \multicolumn{1}{c|}{\textbf{LD}} & \multicolumn{1}{c|}{\textbf{CT}} & \multicolumn{1}{c|}{\textbf{LD}} & \multicolumn{1}{c|}{\textbf{CT}} & \multicolumn{1}{c|}{\textbf{LD}} & \multicolumn{1}{c|}{\textbf{CT}} & \multicolumn{1}{c|}{\textbf{LD}} & \multicolumn{1}{c|}{\textbf{CT}} & \multicolumn{1}{c|}{\textbf{LD}} \\ \hline
		\multicolumn{1}{|c|}{R1}          & \multicolumn{1}{c|}{8}           & \multicolumn{1}{c|}{R24}         & \multicolumn{1}{c|}{13}          & \multicolumn{1}{c|}{R46}         & \multicolumn{1}{c|}{15}          & \multicolumn{1}{c|}{R68}         & \multicolumn{1}{c|}{11}          & \multicolumn{1}{c|}{R90}         & \multicolumn{1}{c|}{18}          & \multicolumn{1}{c|}{C6}          & \multicolumn{1}{c|}{23}          & \multicolumn{1}{c|}{C28}         & \multicolumn{1}{c|}{33}          & \multicolumn{1}{c|}{I2}          & \multicolumn{1}{c|}{50}          \\ \hline
		\multicolumn{1}{|c|}{R2}          & \multicolumn{1}{c|}{9}           & \multicolumn{1}{c|}{R25}         & \multicolumn{1}{c|}{13}          & \multicolumn{1}{c|}{R47}         & \multicolumn{1}{c|}{15}          & \multicolumn{1}{c|}{R69}         & \multicolumn{1}{c|}{13}          & \multicolumn{1}{c|}{R91}         & \multicolumn{1}{c|}{9}           & \multicolumn{1}{c|}{C7}          & \multicolumn{1}{c|}{19}          & \multicolumn{1}{c|}{C29}         & \multicolumn{1}{c|}{21}          & \multicolumn{1}{c|}{I3}          & \multicolumn{1}{c|}{46}          \\ \hline
		\multicolumn{1}{|c|}{R3}          & \multicolumn{1}{c|}{17}          & \multicolumn{1}{c|}{R26}         & \multicolumn{1}{c|}{12}          & \multicolumn{1}{c|}{R48}         & \multicolumn{1}{c|}{7}           & \multicolumn{1}{c|}{R70}         & \multicolumn{1}{c|}{18}          & \multicolumn{1}{c|}{R92}         & \multicolumn{1}{c|}{16}          & \multicolumn{1}{c|}{C8}          & \multicolumn{1}{c|}{25}          & \multicolumn{1}{c|}{C30}         & \multicolumn{1}{c|}{41}          & \multicolumn{1}{c|}{I4}          & \multicolumn{1}{c|}{54}          \\ \hline
		\multicolumn{1}{|c|}{R4}          & \multicolumn{1}{c|}{13}          & \multicolumn{1}{c|}{R27}         & \multicolumn{1}{c|}{18}          & \multicolumn{1}{c|}{R49}         & \multicolumn{1}{c|}{11}          & \multicolumn{1}{c|}{R71}         & \multicolumn{1}{c|}{8}           & \multicolumn{1}{c|}{R93}         & \multicolumn{1}{c|}{9}           & \multicolumn{1}{c|}{C9}          & \multicolumn{1}{c|}{20}          & \multicolumn{1}{c|}{C31}         & \multicolumn{1}{c|}{36}          & \multicolumn{1}{c|}{I5}          & \multicolumn{1}{c|}{65}          \\ \hline
		\multicolumn{1}{|c|}{R5}          & \multicolumn{1}{c|}{12}          & \multicolumn{1}{c|}{R28}         & \multicolumn{1}{c|}{11}          & \multicolumn{1}{c|}{R50}         & \multicolumn{1}{c|}{12}          & \multicolumn{1}{c|}{R72}         & \multicolumn{1}{c|}{10}          & \multicolumn{1}{c|}{R94}         & \multicolumn{1}{c|}{10}          & \multicolumn{1}{c|}{C10}         & \multicolumn{1}{c|}{26}          & \multicolumn{1}{c|}{C32}         & \multicolumn{1}{c|}{17}          & \multicolumn{1}{c|}{I6}          & \multicolumn{1}{c|}{75}          \\ \hline
		\multicolumn{1}{|c|}{R6}          & \multicolumn{1}{c|}{13}          & \multicolumn{1}{c|}{R29}         & \multicolumn{1}{c|}{11}          & \multicolumn{1}{c|}{R51}         & \multicolumn{1}{c|}{12}          & \multicolumn{1}{c|}{R73}         & \multicolumn{1}{c|}{9}           & \multicolumn{1}{c|}{R95}         & \multicolumn{1}{c|}{16}          & \multicolumn{1}{c|}{C11}         & \multicolumn{1}{c|}{30}          & \multicolumn{1}{c|}{C33}         & \multicolumn{1}{c|}{28}          & \multicolumn{1}{c|}{I7}          & \multicolumn{1}{c|}{88}          \\ \hline
		\multicolumn{1}{|c|}{R7}          & \multicolumn{1}{c|}{16}          & \multicolumn{1}{c|}{R30}         & \multicolumn{1}{c|}{8}           & \multicolumn{1}{c|}{R52}         & \multicolumn{1}{c|}{14}          & \multicolumn{1}{c|}{R74}         & \multicolumn{1}{c|}{8}           & \multicolumn{1}{c|}{R96}         & \multicolumn{1}{c|}{12}          & \multicolumn{1}{c|}{C12}         & \multicolumn{1}{c|}{15}          & \multicolumn{1}{c|}{C34}         & \multicolumn{1}{c|}{36}          & \multicolumn{1}{c|}{I8}          & \multicolumn{1}{c|}{92}          \\ \hline
		\multicolumn{1}{|c|}{R8}          & \multicolumn{1}{c|}{12}          & \multicolumn{1}{c|}{R31}         & \multicolumn{1}{c|}{9}           & \multicolumn{1}{c|}{R53}         & \multicolumn{1}{c|}{17}          & \multicolumn{1}{c|}{R75}         & \multicolumn{1}{c|}{15}          & \multicolumn{1}{c|}{R97}         & \multicolumn{1}{c|}{9}           & \multicolumn{1}{c|}{C13}         & \multicolumn{1}{c|}{15}          & \multicolumn{1}{c|}{C35}         & \multicolumn{1}{c|}{17}          & \multicolumn{1}{c|}{I9}          & \multicolumn{1}{c|}{30}          \\ \hline
		\multicolumn{1}{|c|}{R9}          & \multicolumn{1}{c|}{11}          & \multicolumn{1}{c|}{R32}         & \multicolumn{1}{c|}{16}          & \multicolumn{1}{c|}{R54}         & \multicolumn{1}{c|}{17}          & \multicolumn{1}{c|}{R76}         & \multicolumn{1}{c|}{15}          & \multicolumn{1}{c|}{R98}         & \multicolumn{1}{c|}{15}          & \multicolumn{1}{c|}{C14}         & \multicolumn{1}{c|}{15}          & \multicolumn{1}{c|}{C36}         & \multicolumn{1}{c|}{17}          & \multicolumn{1}{c|}{I10}         & \multicolumn{1}{c|}{42}          \\ \hline
		\multicolumn{1}{|c|}{R10}         & \multicolumn{1}{c|}{13}          & \multicolumn{1}{c|}{R33}         & \multicolumn{1}{c|}{7}           & \multicolumn{1}{c|}{R55}         & \multicolumn{1}{c|}{8}           & \multicolumn{1}{c|}{R77}         & \multicolumn{1}{c|}{13}          & \multicolumn{1}{c|}{R99}         & \multicolumn{1}{c|}{8}           & \multicolumn{1}{c|}{C15}         & \multicolumn{1}{c|}{15}          & \multicolumn{1}{c|}{C37}         & \multicolumn{1}{c|}{35}          & \multicolumn{1}{c|}{I11}         & \multicolumn{1}{c|}{54}          \\ \hline
		\multicolumn{1}{|c|}{R11}         & \multicolumn{1}{c|}{15}          & \multicolumn{1}{c|}{R34}         & \multicolumn{1}{c|}{12}          & \multicolumn{1}{c|}{R56}         & \multicolumn{1}{c|}{11}          & \multicolumn{1}{c|}{R78}         & \multicolumn{1}{c|}{11}          & \multicolumn{1}{c|}{R100}        & \multicolumn{1}{c|}{17}          & \multicolumn{1}{c|}{C16}         & \multicolumn{1}{c|}{30}          & \multicolumn{1}{c|}{C38}         & \multicolumn{1}{c|}{15}          & \multicolumn{1}{c|}{I12}         & \multicolumn{1}{c|}{62}          \\ \hline
		\multicolumn{1}{|c|}{R12}         & \multicolumn{1}{c|}{13}          & \multicolumn{1}{c|}{R35}         & \multicolumn{1}{c|}{16}          & \multicolumn{1}{c|}{R57}         & \multicolumn{1}{c|}{17}          & \multicolumn{1}{c|}{R79}         & \multicolumn{1}{c|}{12}          & \multicolumn{1}{c|}{R101}        & \multicolumn{1}{c|}{8}           & \multicolumn{1}{c|}{C17}         & \multicolumn{1}{c|}{35}          & \multicolumn{1}{c|}{C39}         & \multicolumn{1}{c|}{25}          & \multicolumn{1}{c|}{I13}         & \multicolumn{1}{c|}{68}          \\ \hline
		\multicolumn{1}{|c|}{R13}         & \multicolumn{1}{c|}{13}          & \multicolumn{1}{c|}{R36}         & \multicolumn{1}{c|}{9}           & \multicolumn{1}{c|}{R58}         & \multicolumn{1}{c|}{9}           & \multicolumn{1}{c|}{R80}         & \multicolumn{1}{c|}{10}          & \multicolumn{1}{c|}{R102}        & \multicolumn{1}{c|}{14}          & \multicolumn{1}{c|}{C18}         & \multicolumn{1}{c|}{25}          & \multicolumn{1}{c|}{C40}         & \multicolumn{1}{c|}{28}          & \multicolumn{1}{c|}{I14}         & \multicolumn{1}{c|}{74}          \\ \hline
		\multicolumn{1}{|c|}{R14}         & \multicolumn{1}{c|}{13}          & \multicolumn{1}{c|}{R37}         & \multicolumn{1}{c|}{10}          & \multicolumn{1}{c|}{R59}         & \multicolumn{1}{c|}{18}          & \multicolumn{1}{c|}{R81}         & \multicolumn{1}{c|}{12}          & \multicolumn{1}{c|}{R103}        & \multicolumn{1}{c|}{8}           & \multicolumn{1}{c|}{C19}         & \multicolumn{1}{c|}{18}          & \multicolumn{1}{c|}{C41}         & \multicolumn{1}{c|}{17}          & \multicolumn{1}{c|}{I15}         & \multicolumn{1}{c|}{90}          \\ \hline
		\multicolumn{1}{|c|}{R15}         & \multicolumn{1}{c|}{12}          & \multicolumn{1}{c|}{R38}         & \multicolumn{1}{c|}{18}          & \multicolumn{1}{c|}{R60}         & \multicolumn{1}{c|}{14}          & \multicolumn{1}{c|}{R82}         & \multicolumn{1}{c|}{11}          & \multicolumn{1}{c|}{R104}        & \multicolumn{1}{c|}{13}          & \multicolumn{1}{c|}{C20}         & \multicolumn{1}{c|}{23}          & \multicolumn{1}{c|}{C42}         & \multicolumn{1}{c|}{20}          & \multicolumn{1}{c|}{I16}         & \multicolumn{1}{c|}{30}          \\ \hline
		\multicolumn{1}{|c|}{R16}         & \multicolumn{1}{c|}{7}           & \multicolumn{1}{c|}{R39}         & \multicolumn{1}{c|}{13}          & \multicolumn{1}{c|}{R61}         & \multicolumn{1}{c|}{8}           & \multicolumn{1}{c|}{R83}         & \multicolumn{1}{c|}{13}          & \multicolumn{1}{c|}{R105}        & \multicolumn{1}{c|}{8}           & \multicolumn{1}{c|}{C21}         & \multicolumn{1}{c|}{33}          & \multicolumn{1}{c|}{C43}         & \multicolumn{1}{c|}{24}          & \multicolumn{1}{c|}{I17}         & \multicolumn{1}{c|}{30}          \\ \hline
		\multicolumn{1}{|c|}{R17}         & \multicolumn{1}{c|}{16}          & \multicolumn{1}{c|}{R40}         & \multicolumn{1}{c|}{15}          & \multicolumn{1}{c|}{R62}         & \multicolumn{1}{c|}{16}          & \multicolumn{1}{c|}{R84}         & \multicolumn{1}{c|}{13}          & \multicolumn{1}{c|}{R106}        & \multicolumn{1}{c|}{17}          & \multicolumn{1}{c|}{C22}         & \multicolumn{1}{c|}{16}          & \multicolumn{1}{c|}{C44}         & \multicolumn{1}{c|}{28}          & \multicolumn{1}{c|}{I18}         & \multicolumn{1}{c|}{60}          \\ \hline
		\multicolumn{1}{|c|}{R18}         & \multicolumn{1}{c|}{16}          & \multicolumn{1}{c|}{R41}         & \multicolumn{1}{c|}{16}          & \multicolumn{1}{c|}{R63}         & \multicolumn{1}{c|}{14}          & \multicolumn{1}{c|}{R85}         & \multicolumn{1}{c|}{16}          & \multicolumn{1}{c|}{C1}          & \multicolumn{1}{c|}{25}          & \multicolumn{1}{c|}{C23}         & \multicolumn{1}{c|}{45}          & \multicolumn{1}{c|}{C45}         & \multicolumn{1}{c|}{31}          & \multicolumn{1}{c|}{I19}         & \multicolumn{1}{c|}{25}          \\ \hline
		\multicolumn{1}{|c|}{R19}         & \multicolumn{1}{c|}{17}          & \multicolumn{1}{c|}{R42}         & \multicolumn{1}{c|}{8}           & \multicolumn{1}{c|}{R64}         & \multicolumn{1}{c|}{13}          & \multicolumn{1}{c|}{R86}         & \multicolumn{1}{c|}{7}           & \multicolumn{1}{c|}{C2}          & \multicolumn{1}{c|}{15}          & \multicolumn{1}{c|}{C24}         & \multicolumn{1}{c|}{45}          & \multicolumn{1}{c|}{C46}         & \multicolumn{1}{c|}{22}          & \multicolumn{1}{c|}{I20}         & \multicolumn{1}{c|}{35}          \\ \hline
		\multicolumn{1}{|c|}{R20}         & \multicolumn{1}{c|}{7}           & \multicolumn{1}{c|}{R43}         & \multicolumn{1}{c|}{17}          & \multicolumn{1}{c|}{R65}         & \multicolumn{1}{c|}{9}           & \multicolumn{1}{c|}{R87}         & \multicolumn{1}{c|}{6}           & \multicolumn{1}{c|}{C3}          & \multicolumn{1}{c|}{20}          & \multicolumn{1}{c|}{C25}         & \multicolumn{1}{c|}{19}          & \multicolumn{1}{c|}{C47}         & \multicolumn{1}{c|}{32}          & \multicolumn{1}{c|}{I21}         & \multicolumn{1}{c|}{40}          \\ \hline
		\multicolumn{1}{|c|}{R21}         & \multicolumn{1}{c|}{12}          & \multicolumn{1}{c|}{R44}         & \multicolumn{1}{c|}{7}           & \multicolumn{1}{c|}{R66}         & \multicolumn{1}{c|}{6}           & \multicolumn{1}{c|}{R88}         & \multicolumn{1}{c|}{15}          & \multicolumn{1}{c|}{C4}          & \multicolumn{1}{c|}{19}          & \multicolumn{1}{c|}{C26}         & \multicolumn{1}{c|}{25}          & \multicolumn{1}{c|}{C48}         & \multicolumn{1}{c|}{28}          & \multicolumn{1}{c|}{I22}         & \multicolumn{1}{c|}{38}          \\ \hline
		\multicolumn{1}{|c|}{R22}         & \multicolumn{1}{c|}{11}          & \multicolumn{1}{c|}{R45}         & \multicolumn{1}{c|}{15}          & \multicolumn{1}{c|}{R67}         & \multicolumn{1}{c|}{9}           & \multicolumn{1}{c|}{R89}         & \multicolumn{1}{c|}{12}          & \multicolumn{1}{c|}{C5}          & \multicolumn{1}{c|}{18}          & \multicolumn{1}{c|}{C27}         & \multicolumn{1}{c|}{25}          & \multicolumn{1}{c|}{I1}          & \multicolumn{1}{c|}{40}          & \multicolumn{1}{c|}{I23}         & \multicolumn{1}{c|}{42}          \\ \hline
		\multicolumn{1}{|c|}{R23}         & \multicolumn{1}{c|}{8}           & \multicolumn{1}{c|}{}            & \multicolumn{1}{c|}{}            & \multicolumn{1}{c|}{}            & \multicolumn{1}{c|}{}            & \multicolumn{1}{c|}{}            & \multicolumn{1}{c|}{}            & \multicolumn{1}{c|}{}            & \multicolumn{1}{c|}{}            & \multicolumn{1}{c|}{}            & \multicolumn{1}{c|}{}            & \multicolumn{1}{c|}{}            & \multicolumn{1}{c|}{}            & \multicolumn{1}{c|}{}            & \multicolumn{1}{c|}{}            \\ \hline
		\multicolumn{16}{l}{\tiny{\textbf{CT: customer type (R: residential, C: commercial, I: industrial); LD: load demand}}}                                                                                                                                                                                                                                                                                                                                                                                                                                                                        
	\end{tabular}
\end{table*}

\ifCLASSOPTIONcaptionsoff
  \newpage
\fi

\bibliographystyle{IEEEtran}
\bibliography{IEEEreferences} 

% Generated by IEEEtran.bst, version: 1.12 (2007/01/11)
\begin{thebibliography}{10}
\providecommand{\url}[1]{#1}
\csname url@samestyle\endcsname
\providecommand{\newblock}{\relax}
\providecommand{\bibinfo}[2]{#2}
\providecommand{\BIBentrySTDinterwordspacing}{\spaceskip=0pt\relax}
\providecommand{\BIBentryALTinterwordstretchfactor}{4}
\providecommand{\BIBentryALTinterwordspacing}{\spaceskip=\fontdimen2\font plus
\BIBentryALTinterwordstretchfactor\fontdimen3\font minus
  \fontdimen4\font\relax}
\providecommand{\BIBforeignlanguage}[2]{{%
\expandafter\ifx\csname l@#1\endcsname\relax
\typeout{** WARNING: IEEEtran.bst: No hyphenation pattern has been}%
\typeout{** loaded for the language `#1'. Using the pattern for}%
\typeout{** the default language instead.}%
\else
\language=\csname l@#1\endcsname
\fi
#2}}
\providecommand{\BIBdecl}{\relax}
\BIBdecl

\bibitem{faruqui2010impact}
A.~Faruqui, S.~Sergici, and J.~Palmer, ``The impact of dynamic pricing on low
  income customers,'' \emph{Institute for Electric Efficiency Whitepaper},
  2010.

\bibitem{1198281}
D.~S. {Kirschen}, ``Demand-side view of electricity markets,'' \emph{IEEE
  Transactions on Power Systems}, vol.~18, no.~2, pp. 520--527, May 2003.

\bibitem{doe2006benefits}
U.~DoE, ``Benefits of demand response in electricity markets and
  recommendations for achieving them. a report to the united states congress
  pursuant to section 1252 of the energy policy act of 2005,'' in \emph{US
  Washington, DC: Department of Energy.[http://eetd. lbl.
  gov/ea/EMP/reports/congress-1252d. pdf](26 July 2009)}, 2006.

\bibitem{staff2006assessment}
F.~Staff, ``Assessment of demand response and advanced metering,''
  \emph{Federal Energy Regulatory Commission, Docket AD-06-2-000}, 2006.

\bibitem{federal2007staff}
F.~E.~R. Commission \emph{et~al.}, ``Staff report-assessment of demand response
  and advanced metering,'' 2007.

\bibitem{us2009national}
U.~F. E.~R. Commission \emph{et~al.}, ``A national assessment of demand
  response potential,'' Staff Report, Washington, DC. p, Tech. Rep., 2009.

\bibitem{yang2017opportunities}
C.-J. Yang, ``Opportunities and barriers to demand response in china,''
  \emph{Resources, Conservation and Recycling}, vol. 121, pp. 51--55, 2017.

\bibitem{article}
C.~Goldman, M.~Reid, R.~Levy, and A.~Silverstein, ``Coordination of energy
  efficiency and demand response,'' 06 2010.

\bibitem{ahn2011optimal}
C.~Ahn, C.-T. Li, and H.~Peng, ``Optimal decentralized charging control
  algorithm for electrified vehicles connected to smart grid,'' \emph{Journal
  of Power Sources}, vol. 196, no.~23, pp. 10\,369--10\,379, 2011.

\bibitem{kirschen2005fundamentals}
D.~Kirschen and G.~Strbac, ``Fundamentals of power system economics:
  Copyright{\copyright} 2004 john wiley \& sons,'' 2005.

\bibitem{hussain2018review}
M.~Hussain and Y.~Gao, ``A review of demand response in an efficient smart grid
  environment,'' \emph{The Electricity Journal}, vol.~31, no.~5, pp. 55--63,
  2018.

\bibitem{6254817}
Q.~{Zhang} and J.~{Li}, ``Demand response in electricity markets: A review,''
  in \emph{2012 9th International Conference on the European Energy Market},
  May 2012, pp. 1--8.

\bibitem{ghazvini2017demand}
M.~A.~F. Ghazvini, J.~Soares, O.~Abrishambaf, R.~Castro, and Z.~Vale, ``Demand
  response implementation in smart households,'' \emph{Energy and buildings},
  vol. 143, pp. 129--148, 2017.

\bibitem{palensky2011demand}
P.~Palensky and D.~Dietrich, ``Demand side management: Demand response,
  intelligent energy systems, and smart loads,'' \emph{IEEE transactions on
  industrial informatics}, vol.~7, no.~3, pp. 381--388, 2011.

\bibitem{32577}
B.~{Daryanian}, R.~E. {Bohn}, and R.~D. {Tabors}, ``Optimal demand-side
  response to electricity spot prices for storage-type customers,'' \emph{IEEE
  Transactions on Power Systems}, vol.~4, no.~3, pp. 897--903, Aug 1989.

\bibitem{651628}
J.~G. {Roos} and I.~E. {Lane}, ``Industrial power demand response analysis for
  one-part real-time pricing,'' \emph{IEEE Transactions on Power Systems},
  vol.~13, no.~1, pp. 159--164, Feb 1998.

\bibitem{schwarz2002industrial}
P.~M. Schwarz, T.~N. Taylor, M.~Birmingham, and S.~L. Dardan, ``Industrial
  response to electricity real-time prices: Short run and long run,''
  \emph{Economic Inquiry}, vol.~40, no.~4, pp. 597--610, 2002.

\bibitem{faruqui2012time}
A.~Faruqui, R.~Hledik, and J.~Palmer, \emph{Time-varying and dynamic rate
  design}.\hskip 1em plus 0.5em minus 0.4em\relax Regulatory Assistance
  Project, 2012.

\bibitem{wenders1976peak}
J.~T. Wenders, ``Peak load pricing in the electric utility industry,''
  \emph{The Bell Journal of Economics}, pp. 232--241, 1976.

\bibitem{sheen1995response}
J.-N. Sheen, C.-S. Chen, and T.-Y. Wang, ``Response of large industrial
  customers to electricity pricing by voluntary time-of-use in taiwan,''
  \emph{IEE Proceedings-Generation, Transmission and Distribution}, vol. 142,
  no.~2, pp. 157--166, 1995.

\bibitem{yu2006optimal}
N.~Yu and J.-l. Yu, ``Optimal tou decision considering demand response model,''
  in \emph{2006 International Conference on Power System Technology}.\hskip 1em
  plus 0.5em minus 0.4em\relax IEEE, 2006, pp. 1--5.

\bibitem{david1993effect}
A.~David and Y.~Li, ``Effect of inter-temporal factors on the real time pricing
  of electricity,'' \emph{IEEE transactions on power systems}, vol.~8, no.~1,
  pp. 44--52, 1993.

\bibitem{kirschen2000factoring}
D.~S. Kirschen, G.~Strbac, P.~Cumperayot, and D.~de~Paiva~Mendes, ``Factoring
  the elasticity of demand in electricity prices,'' \emph{IEEE Transactions on
  Power Systems}, vol.~15, no.~2, pp. 612--617, 2000.

\bibitem{celebi2007model}
E.~Celebi and J.~D. Fuller, ``A model for efficient consumer pricing schemes in
  electricity markets,'' \emph{IEEE Transactions on Power Systems}, vol.~22,
  no.~1, pp. 60--67, 2007.

\bibitem{sheen1994time}
J.-N. Sheen, C.-S. Chen, and J.-K. Yang, ``Time-of-use pricing for load
  management programs in taiwan power company,'' \emph{IEEE Transactions on
  Power Systems}, vol.~9, no.~1, pp. 388--396, 1994.

\bibitem{aalami2008demand}
H.~Aalami, G.~Yousefi, and M.~P. Moghadam, ``Demand response model considering
  edrp and tou programs,'' in \emph{2008 IEEE/PES Transmission and Distribution
  Conference and Exposition}.\hskip 1em plus 0.5em minus 0.4em\relax IEEE,
  2008, pp. 1--6.

\bibitem{duan2004risk}
D.~Duan, J.~Liu, H.~Niu, and J.~Wu, ``A risk-evasion tou pricing method for
  distribution utility in deregulated market environment,'' in \emph{2004
  International Conference on Power System Technology, 2004. PowerCon 2004.},
  vol.~1.\hskip 1em plus 0.5em minus 0.4em\relax IEEE, 2004, pp. 527--531.

\bibitem{FARUQUI201261}
\BIBentryALTinterwordspacing
A.~Faruqui, ``Chapter 3 - the ethics of dynamic pricing,'' in \emph{Smart
  Grid}, F.~P. Sioshansi, Ed.\hskip 1em plus 0.5em minus 0.4em\relax Boston:
  Academic Press, 2012, pp. 61 -- 83. [Online]. Available:
  \url{http://www.sciencedirect.com/science/article/pii/B9780123864529000036}
\BIBentrySTDinterwordspacing

\bibitem{5677457}
D.~T. {Nguyen}, M.~{Negnevitsky}, and M.~{de Groot}, ``Pool-based demand
  response exchange—concept and modeling,'' \emph{IEEE Transactions on Power
  Systems}, vol.~26, no.~3, pp. 1677--1685, Aug 2011.

\bibitem{6345351}
W.~{Zhang}, K.~{Kalsi}, J.~{Fuller}, M.~{Elizondo}, and D.~{Chassin},
  ``Aggregate model for heterogeneous thermostatically controlled loads with
  demand response,'' in \emph{2012 IEEE Power and Energy Society General
  Meeting}, July 2012, pp. 1--8.

\bibitem{7972993}
H.~{Bitaraf} and S.~{Rahman}, ``Reducing curtailed wind energy through energy
  storage and demand response,'' \emph{IEEE Transactions on Sustainable
  Energy}, vol.~9, no.~1, pp. 228--236, Jan 2018.

\bibitem{allcott2011rethinking}
H.~Allcott, ``Rethinking real-time electricity pricing,'' \emph{Resource and
  energy economics}, vol.~33, no.~4, pp. 820--842, 2011.

\bibitem{ericson2009direct}
T.~Ericson, ``Direct load control of residential water heaters,'' \emph{Energy
  Policy}, vol.~37, no.~9, pp. 3502--3512, 2009.

\bibitem{berg1998basics}
S.~Berg, ``Basics of rate design: Pricing principles and self-selecting
  two-part tariffs,'' \emph{Infrastructure regulation and market reform:
  Principles and practice}, pp. 74--90, 1998.

\bibitem{lin2013electricity}
B.~Lin and X.~Liu, ``Electricity tariff reform and rebound effect of
  residential electricity consumption in china,'' \emph{Energy}, vol.~59, pp.
  240--247, 2013.

\bibitem{cutter2012maximizing}
E.~Cutter, C.~K. Woo, F.~Kahrl, and A.~Taylor, ``Maximizing the value of
  responsive load,'' \emph{The Electricity Journal}, vol.~25, no.~7, pp. 6--16,
  2012.

\bibitem{star2010dynamic}
A.~Star, M.~Isaacson, D.~Haeg, L.~Kotewa, and C.~Energy, ``The dynamic pricing
  mousetrap: Why isn’t the world beating down our door,'' in \emph{ACEEE
  summer study on energy efficiency in buildings}, 2010, pp. 257--268.

\bibitem{5156266}
C.~{Su} and D.~{Kirschen}, ``Quantifying the effect of demand response on
  electricity markets,'' \emph{IEEE Transactions on Power Systems}, vol.~24,
  no.~3, pp. 1199--1207, Aug 2009.

\bibitem{brubaker1975free}
E.~R. Brubaker, ``Free ride, free revelation, or golden rule?'' \emph{The
  Journal of Law and Economics}, vol.~18, no.~1, pp. 147--161, 1975.

\bibitem{bagnoli1989provision}
M.~Bagnoli and B.~L. Lipman, ``Provision of public goods: Fully implementing
  the core through private contributions,'' \emph{The Review of Economic
  Studies}, vol.~56, no.~4, pp. 583--601, 1989.

\bibitem{baran1989network}
M.~E. Baran and F.~F. Wu, ``Network reconfiguration in distribution systems for
  loss reduction and load balancing,'' \emph{IEEE Power Engineering Review},
  vol.~9, no.~4, pp. 101--102, 1989.

\bibitem{IEX}
Indian Energy Exchange [Online] Available: \url{https://www.iexindia.com/ }.

\bibitem{8272222}
N.~{Kanwar}, N.~{Gupta}, K.~R. {Niazi}, and A.~{Swarnkar}, ``Optimal
  distributed resource planning for microgrids under uncertain environment,''
  \emph{IET Renewable Power Generation}, vol.~12, no.~2, pp. 244--251, 2018.

\bibitem{faruqui2011dynamic}
A.~Faruqui and J.~Palmer, ``Dynamic pricing and its discontents,''
  \emph{Regulation}, vol.~34, p.~16, 2011.

\end{thebibliography}
\end{document}